\documentclass[twocolumn]{revtex4}

\usepackage[dvips]{graphicx}
\usepackage{amssymb,amsfonts,amsmath}

\begin{document}

\title{Observation of a multiferroic critical end point}

\author{Jae Wook Kim$^{1}$, S. Y. Haam$^{1}$, Y. S. Oh$^{1}$,
S.~Park$^{2}$, S.-W.~Cheong$^{2}$, P.~A.~Sharma$^{3}$,
M.~Jaime$^{3}$, N.~Harrison$^{3}$, Jung~Hoon~Han$^{4}$, Gun-Sang
Jeon$^{1}$, P.~Coleman$^{2}$, and Kee Hoon Kim$^{1*}$}

\address{$^{1}$CSCMR \& FPRD, Department of Physics and
Astronomy, Seoul National University, Seoul 151-742, South Korea.\\
$^{2}$Rutgers Center for Emergent Materials and Department of
Physics and Astronomy, Rutgers University,
Piscataway, NJ 08854, USA.\\
$^{3}$National High Magnetic Field Laboratory,
LANL, Los Alamos, NM 87545, USA.\\
$^{4}$Department of Physics and BK21 Physics Research Division,
Sungkyunkwan University, Suwon 440-746, South Korea.}

\maketitle \textbf{The study of abrupt increases in magnetization
with magnetic field known as metamagnetic transitions has opened a
rich vein of new physics in itinerant electron systems, including
the discovery of quantum critical end points with a marked
propensity to develop new kinds of order. However, the electric
analogue of the metamagnetic critical end point, a
``metaelectric'' critical end point has not yet been realized.
Multiferroic materials wherein magnetism and ferroelectricity are
cross-coupled are ideal candidates for the exploration of this
novel possibility using magnetic-field (\emph{H}) as a tuning
parameter. Herein, we report the discovery of a
magnetic-field-induced metaelectric transition in multiferroic
BiMn$_{2}$O$_{5}$ in which the electric polarization (\emph{P})
switches polarity along with a concomitant Mn spin-flop transition
at a critical magnetic field \emph{H}$_{\rm c}$. The simultaneous
metaelectric and spin-flop transitions become sharper upon
cooling, but remain a continuous crossover even down to 0.5 K.
Near the \emph{P}=0 line realized at $\mu_{0}$\emph{H}$_{\rm
c}$$\approx$18 T below 20 K, the dielectric constant
($\varepsilon$) increases significantly over wide field- and
temperature (\emph{T})-ranges. Furthermore, a characteristic
power-law behavior is found in the \emph{P}(\emph{H}) and
$\varepsilon$(\emph{H}) curves at \emph{T}=0.66 K. These findings
indicate that a magnetic-field-induced metaelectric critical end
point is realized in BiMn$_2$O$_5$ near zero temperature.}

The term ``critical end point'' refers to a singular point in the
phase diagram of matter at the end of a 1st order phase line, as
for example, the liquid-gas critical point of water. The
importance of this special point for broad classes of matter has
rapidly increased in recent years \cite{grigera1,neil}. Not only
can it provide large thermal fluctuations as a necessary
ingredient for displaying universal power-law of physical
quantities, but in the special case where the phase transition is
suppressed to zero temperature, it can give rise to intense
quantum mechanical fluctuations with a marked propensity to
develop instabilities into new kinds of ground state. The latter
case has been recently realized in itinerant metamagnets such as
Sr$_3$Ru$_2$O$_7$ \cite{grigera2} and URu$_2$Si$_2$ \cite{khkim1}.
These developments motivate a parallel search for a metaelectric
critical end point. By analogy with magnetism, one might expect
such sudden changes in electric polarization to be a rather
general phenomenon in (anti)ferroelectric materials under
application of an electric field ($E$). However, to date, those
effects have been limitedly observed in specific systems such as
relaxor ferroelectrics \cite{Kutnjak} and DyVO$_4$ with the
Jahn-Teller structural distortion \cite{vekhter}. One possible
reason for the scarcity of the phenomenon is the practical
difficulty of applying the large voltages required ($\gtrapprox$1
kV) without inducing electrical breakdown. Magnetic fields may in
fact be better candidates for inducing metaelectric transitions,
for not only do they avoid the problem of electrical breakdown,
they provide a reversible method of tuning matter with many
well-established advantages over alternative methods such as
pressure or chemical doping. This is still a challenging task
because of the intrinsically small cross-coupling between spin and
lattice degrees of freedom. However, in a special class of matter
called multiferroics where the cross-coupling between electricity
and magnetism is enhanced,
\cite{kimura,hur1,spaldin,eerenstein,tokura,kenzelmann}, such a
rare metaelectric transition, and as its critical end point can be
induced by application of magnetic field, as we show in this
study.

\begin{figure}
\centerline{\includegraphics[width=.48\textwidth]{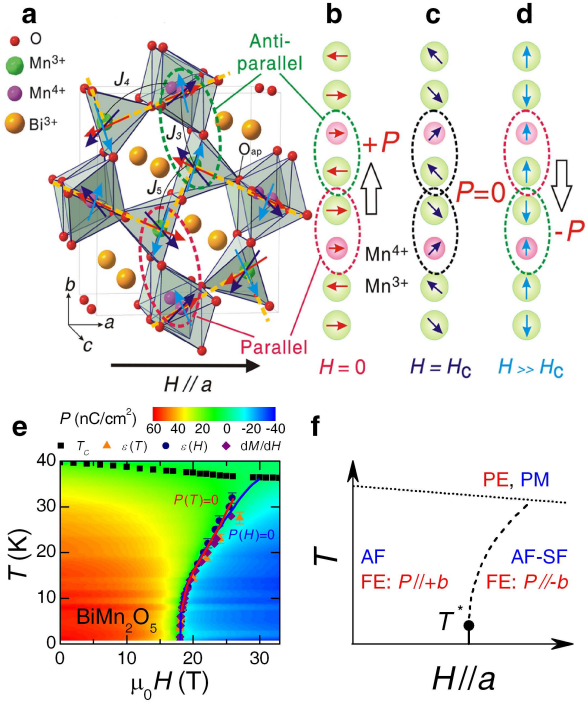}}
\caption{Magnetic structure and phase diagram. (\textit{a}) The
structure and AF zigzag spin chains under various magnetic fields
for BiMn$_{2}$O$_{5}$. \textit{J}$_{\rm 3}$ and \textit{J}$_{\rm
4}$ denote the interchain and intrachain exchange interactions
between Mn$^{3+}$-Mn$^{4+}$ spin pairs, respectively, while
\textit{J}$_{\rm 5}$ describes that between neighboring Mn$^{3+}$
spins in the bi-pyramid. Red, dark blue, and light blue arrows represent
spin directions at \textit{H}=0, \textit{H}=\textit{H}$_{\rm c}$,
and \textit{H}$\gg$\textit{H}$_{\rm c}$, respectively. The
corresponding simplified spin arrangements and the net \textit{P}
(black open arrows) are shown for (\textit{b}) \textit{H}=0,
(\textit{c}) \textit{H}=\textit{H}$_{\rm c}$, and (\textit{d})
\textit{H}$\gg$\textit{H}$_{\rm c}$. (\textit{e}) The
\textit{H}-\textit{T} phase diagram and the intensity contour of
\textit{P} based on the \textit{P}(\textit{H}) data. All symbols
were determined from the peak positions of
$\varepsilon$(\textit{H}), $\varepsilon$(\textit{H}), and
\textit{dM/dH}. (\textit{f}) Suggested schematic phase diagram.
Dotted, solid, and dashed lines indicate the 2nd-order
paraelectric (PE)-FE transition, 1st-order AF to spin-flopped AF
(AF-SF) transition, and AF to AF-SF crossover, respectively. A
possible critical end point \textit{T}$^{*}$ is denoted as a solid
circle.}\label{fig1}
\end{figure}

Among recently discovered multiferroic materials, the family of
$R$Mn$_2$O$_5$ ($R$=rare earth, Y) are of particular interest,
because of their large dielectric response under magnetic fields
\cite{hur1}. While BiMn$_2$O$_5$ is isostructural to
$R$Mn$_2$O$_5$ (Fig. 1\textit{a}), its magnetic ground state is
distinct from those of other $R$Mn$_2$O$_5$ compounds. Only
BiMn$_2$O$_5$ remains in a commensurate antiferromagnetic (AF)
phase below $T_{\rm N}$$\sim$40K with a propagation vector of
\textbf{Q}=(0.5, 0, 0.5) \cite{munoz,vec}, while other
$R$Mn$_2$O$_5$ compounds develop an instability into an
incommensurate (IC) AF structure below $T_{\rm IC}$$\approx$25
K\cite{munoz}. According to DC magnetization ($M$) (Fig.
2\textit{a}) and neutron diffraction data \cite{vec},
BiMn$_2$O$_5$ has a typical AF spin ordering with an easy axis
almost parallel to the $a$-axis without showing any signature of a
spiral ordering. At $T_{\rm N}$, the development of magnetic order
is accompanied by the growth of ferroelectric (FE) polarization
$P$ and a sharp peak in the dielectric constant $\varepsilon$
(Figs. 2\textit{b} and 2\textit{c}).

Multiferroicity in materials with incommensurate spiral spin
ordering is normally accounted for in terms of the ``spin-current
model'' \cite{katsura}. Recent neutron scattering studies of
$R$Mn$_2$O$_5$ ($R$=Tb, Ho, Y, and Bi) \cite{vec,chapon1,chapon2}
have suggested an alternative ``exchange-striction'' scenario for
this commensurate magnetic system. The AF phase of $R$Mn$_2$O$_5$
involves zigzag chains of spins (dashed orange lines in Fig.
1\textit{a}) with a staggered moment lying almost parallel to the
Mn$^{3+}$--\ O$_{\rm ap}$ bond (where O$_{\rm ap}$ is the apical
oxygen in each pyramid). Neutron measurements show that the
staggered moment is tilted away from the $a$-axis by about $\pm$8
$^{\circ}$ at low temperatures \cite{vec}, an effect which is
presumably driven by a large single-ion anisotropy in the pyramid
and strong AF exchange interactions along the chain (i.e., large
$J_{\rm 4}$ and $J_{\rm 5}$$>$0 in Fig. 1\textit{a})
\cite{chapon1}. The effect of stacking AF zigzag chains along the
$b$-axis leads to a five member frustrated Mn spin loop in the
$ab$-plane: Mn$^{4+}$--\ Mn$^{3+}$--\ Mn$^{3+}$--\ Mn$^{4+}$--\
Mn$^{3+}$. Although the nearest-neighbor magnetic coupling in the
loop is antiferromagnetic, the odd number of bonds leads
frustrates the spins, preventing them from being be anti-parallel
on every bond in the loop. Due to the relatively small exchange
interaction between Mn$^{3+}$-Mn$^{4+}$ ions ($J_3$), the system
forms a unique spin ordering pattern \cite{vec}: half of the
Mn$^{3+}$--\ Mn$^{4+}$ spin pairs across neighboring zigzags are
approximately antiparallel, whereas the other half are almost
parallel. Under this spin arrangement, exchange-striction between
the Mn$^{3+}$--\ Mn$^{4+}$ spin pairs shifts ions (mostly
Mn$^{3+}$ inside pyramids) in a way that optimizes the
spin-exchange energy: ions with antiparallel spins are pulled
towards each other (green dashed circle in Fig. 1\textit{a}),
while those with parallel spins move apart (red dashed circle)
\cite{cheong}. This exchange-striction mechanism then results in
canted staggered electric dipoles that are aligned parallel to the
Mn$^{3+}$--\ O$_{\rm ap}$ bond in each bi-pyramid, leading to an
electric polarization $P$ along the $b$-axis (Fig. 1\textit{b}).

The application of high magnetic fields along the $a$-axis reveals
unexpected magnetic and electric anomalies (Fig. 2); a sharp
symmetric peak in the $dM/dH$ curve (Fig. 2\textit{e}) is
accompanied by a sharp increase in $\varepsilon$($H$) and a
polarization $P$($H$) which passes through zero (Fig. 2\textit{d}
and 2\textit{f}). The critical fields for these anomalies increase
with temperature, and more or less coincide. Similarly, the
temperature of the $\varepsilon$($T$) peaks correspond to the
temperature where the polarization $P$($T$)=0 (Figs. 2\textit{b}
and 2\textit{c}). A contour intensity plot of the $P$($H$) curves
(Fig. 1\textit{e}) confirms that the trajectory of the $P$($H$)=0
(blue solid line) closely matches that of the $P$($T$)=0 (red
solid line), and the peak positions of $\varepsilon$($H$),
$dM/dH$, and $\varepsilon$($T$) curves below $\sim$20 K.

We attribute the origin of the magnetic anomaly to a spin-flop
(SF) transition in the AF zigzag spin chains. This is supported by
an observation of a small $M$ value and a zero-crossing of the
linearly extrapolated line at $H$$>$$H_{\rm c}$, indicating the
survival of the AF interaction beyond $H_{\rm c}$ (Fig.
2\textit{e}, inset). While the SF is often manifested as a
1st-order transition, in this case the $M$($H$) curve does not
show a discontinuity down to 0.5 K, as evidenced by a lack of
divergence in the $dM/dH$ curve. This characteristic is also
observed in the electric properties; the $P$($H$) curve at 0.66 K
does not show a discontinuity and the $\varepsilon$($H$) curve
does not show divergence even at $T$=0.66 K except sharpening on
cooling. Furthermore, the $\varepsilon$($H$) curves measured in
very slow $H$- or $T$-sweeps did not show any obvious hysteresis
(See, a supplementary figure, Fig. S1). The lack of a sharp
1st-order transition feature becomes more evident in the $P$($T$)
curves (Fig. 2\textit{c}), in which $P$ gradually changes its
magnitude and sign on cooling. In addition, our preliminary
$T$-dependent specific heat measurements across the $P$=0 line
show no peak structure or discontinuity. These observations
strongly suggest that the SF transition and concurrent electrical
transition occur via a smooth crossover down to temperatures as
low as 0.5 K, during which the spin configuration changes
continuously from the AF- (Fig. 1\textit{b}) to the SF-type (Fig.
1\textit{d}).

\begin{figure}
\begin{center}
\centerline{\includegraphics[width=.48\textwidth]{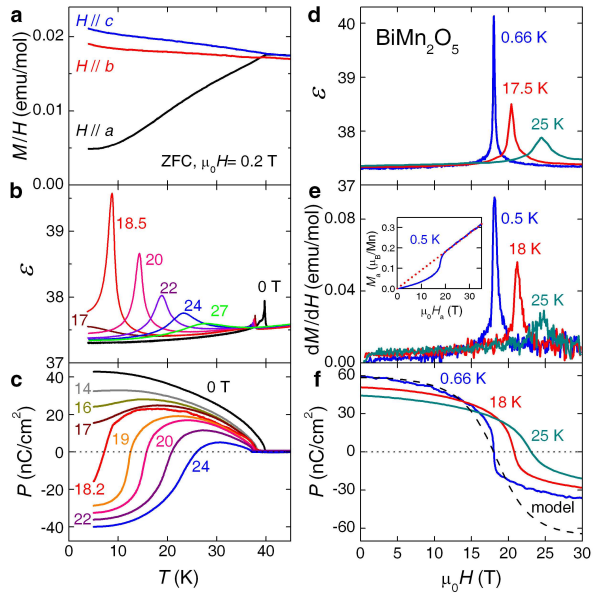}}
\caption{Electric and magnetic properties of BiMn$_{2}$O$_{5}$.
(\textit{a}) \textit{H} measured along the three crystallographic
directions. (\textit{b}) $\varepsilon$(\textit{T}) data down to 4
K at various magnetic fields. The $\varepsilon$(\textit{H}=0) peak
near 40 K represents the FE transition. (\textit{c})
\textit{P}(\textit{T}) curves at various magnetic fields, showing
a crossover from positive to negative. (\textit{d})
$\varepsilon$(\textit{H}) at fixed temperatures. (\textit{e})
Field-dependent \textit{dM/dH} curves at fixed temperatures. The
inset shows the \textit{M}(\textit{H}) curve at \textit{T}=0.5 K.
(\textit{f}) \textit{P}(\textit{H}) at fixed temperatures. The
dashed line shows the calculated \textit{P} curve as predicted by
the spin-rotation model described in the text and in Fig.
1(\textit{b-d}). All dotted lines are guides for the
eye.}\label{fig2}
\end{center}
\end{figure}

In fact, continuous spin-flop transitions are well known in
antiferromagnets when the field $H$ is applied at an angle to the
easy axis: typically, a spin-flip crossover occurs when the angle
between the field and the staggered moment exceeds a certain
critical value\cite{blazey,king}. This case has been well studied
in uniaxial antiferromagnets such as GdAlO$_3$ \cite{blazey} and
MnF$_2$ \cite{king}. In the former, the critical angle starts from
zero at $T_{\rm N}$=3.2 K and grows as the temperature is lowered,
reaching 10$^{\circ}$ at zero temperature. So long as the magnetic
field inclination exceeds the critical angle, a cross-over is
observed, but when the critical angle reaches the
field-inclination angle, the system passes through a critical end
point, located at a temperature $T^*$ in the $H$-$T$ plane and at
lower temperatures the SF sharpens into a first order transition.
In spin flop transitions, the location of $T^{*}$ is determined as
a function of the initial angle between the spin easy axis and the
applied field.

We can easily generalize this understanding to the case of
BiMn$_2$O$_5$. Fig. 1\textit{a} shows that in BiMn$_2$O$_5$, the
zig-zag chain of $Mn$ spins (red arrows) subtend an angle of
$\pm$8 $^{\circ}$ to the a-axis, providing an intrinsic ``tilt''
with respect to a field applied along the $x$ axis. In this way,
the zig-zag chain provides a natural mechanism for a spin-flop
cross-over down to some critical temperature $T^{*}$. This leads
us to identify the $P$=0 trajectory in the phase diagram of
BiMn$_2$O$_5$ (Fig. 1\textit{e}) with the spin-flop crossover line
above $T$$>$$T^*$. Our data enable us to estimate the critical
temperature $T^*$ is at, or below 0.5K.

One of key findings of this work is the SF crossover can drive the
continuous $H$-induced $P$ reversal at $T$$>$$T^*$ via the
exchange-striction mechanism. To understand the mechanism,
consider the effect of a magnetic field $H$ on a single AF zigzag
chain. If we assume that the interchain coupling ($J_3$) can be
neglected, then the dominant intrachain coupling ($J_4$ and $J_5$)
will give rise to an almost antiparallel spin alignment. The
magnetic free energy of the AF zigzag chain at zero temperature
will then be governed by that of a pair of Mn$^{3+}$ spins in the
bi-pyramid, consisting of the energies of the AF spin sublattice
and the single-ion anisotropy of the Mn$^{3+}$ spins
\cite{kanamori}: $F_{\rm
M}$=$-\mu_{0}\{\chi_{\parallel}+(\chi_{\perp}-\chi_{\parallel}){\rm
sin}^{2}\theta\}\emph{H}^{2}$$-\Delta{\rm
cos}^{2}(\theta-\theta_{0})$. Here, $\chi_{\parallel}$ and
$\chi_{\perp}$ are magnetic susceptibility values parallel and
perpendicular to the staggered moment, respectively. The
parameters $\theta$, $\theta$$_0$, and $\Delta$ denote the angle
between the Mn$^{3+}$ spin direction and the $a$-axis, the value
of $\theta$ at zero field and the single-ion anisotropy energy,
respectively. The $\theta$ value at each $H$ leading to the
minimum energy can be obtained from $\partial F_{\rm M}$$/\partial
\theta$=0 and the resultant following relationship:
\begin{equation}\label{eq1}
tan2\theta=\frac{tan2\theta_{0}}{1-\frac{\mu_{0}(\chi_{\perp}-\chi_{\parallel})}{\Delta
cos2\theta_{0}}\emph{H}^2}.
\end{equation}
The above equation predicts that when $\theta_{0}$$>$0 ($<$0),
$\theta$ increases (decreases) continuously with $H$ to cross the
angle 45 $^{\circ}$ (-45 $^{\circ}$) at $H_{\rm c}$=$\sqrt{\Delta
cos2\theta_{0}/\mu_{0}(\chi_{\perp}-\chi_{\parallel})}$. In other
words, the spin rotation occurs in a way that each AF zigzag chain
rotates counter-clockwise or clockwise alternatively to cross 45
$^{\circ}$ or -45 $^{\circ}$ at $H_{\rm c}$ (dark blue arrows in
Fig. 1\textit{a}). When we consider the special $\pm$45 $^{\circ}$
configuration (Fig. 1\textit{c}), in which the angle between
Mn$^{3+}$ and Mn$^{4+}$ spins becomes 90$^{\circ}$, the
exchange-striction is balanced and gives $P$=0. On further
increasing $H$, $|\theta|$ increases to make the spin alignment
almost perpendicular to the field (blue arrows in Fig.
1\textit{a}). As a result, the relative orientation of neighboring
Mn spins reverses as a function of field (see Fig. 1. \textit{b}
and \textit{d}) giving rise to the sign reversal in the electric
polarization $P$.

For comparison with the experimental data, we have calculated
predicted value of
$P$$\propto$$\emph{\textbf{S}}^{4+}$$\cdot$$\emph{\textbf{S}}^{3+}$$\sim$cos2$\theta$
with $\theta$ determined in Eq. (\ref{eq1}). Here,
$\emph{\textbf{S}}^{4+}$ and $\emph{\textbf{S}}^{3+}$ are the spin
vectors of neighboring Mn ions between the AF zigzag chains. The
dashed line in Fig. 2\textit{f} shows the calculated $P$($H$)
curve with $\mu_{0}H_{\rm c}$=18.04 T and $\theta_{0}$=8
$^{\circ}$, showing a qualitative agreement with the curve at
$T$=0.66 K. Therefore, the rather continuous magnetic
field-induced $P$ reversal observed in BiMn$_2$O$_5$ can be
accounted for by the variation of exchange-striction degree driven by
the Mn spin-flop crossover. We further note that only
BiMn$_2$O$_5$ has a homogeneous positive $P$ below $T_{\rm C}$
under small electric poling; all the other $R$Mn$_2$O$_5$
develop additional FE domains with negative polarity and antiphase
domain wall boundaries below $T_{\rm IC}$ \cite{koo}. In turn, the
above $P$ reversal model can be applied to BiMn$_2$O$_5$ only, and
it appears to explain why the experimental finding of $P$ reversal
as well as $\varepsilon$ increase near $H_{\rm c}$ are unique to
BiMn$_2$O$_5$ \cite{haam}.

The magnetically induced polarization reversal we observe is
closely analogous to the field-induced metamagnetic transitions
observed in spin systems \cite{stryj,grigera1} and is most
naturally regarded as a magnetic-field induced ``metaelectric''
transition \cite{vekhter}. As the inversion symmetry is already
broken in both sides of the $H_{\rm c}$($T$) boundary in Fig.
1\textit{f}, the metaelectric transition involves only a change in
the magnitude of the ferroelectric order parameter $P$. In this
case, due to the coupling of $P$ to the Mn spin moment, the
metaelectric transition occurs as a crossover down to very low
temperatures ($T$=0.5 K). The systematically increased sharpening
observed in $\varepsilon$($H$), $dM/dH$ and $P$($H$) curves upon
cooling all indicate that $T^*$ is close to 0.5 K. If so, we would
expect to observe critical fluctuations associated with the
critical end point, with a power-law evolution of the physical
quantities in close proximity to $T^*$. Our experiment confirms
this expectation as shown in Figs. 3\textit{a} and 3\textit{b},
where ($\varepsilon$($H$)-$\varepsilon$($H$=0))$^{-1}$ vs.
$h^{2/3}$ and $|P|$ vs. $h^{1/3}$ plots produce quite linear
behavior. Here $h\equiv|(H$-$H_{\rm c}/ H_{\rm c}|$ is the
normalized deviation from the critical field $H_{\rm c}$=18.04 T.
The linearity observed over a broad range of $h$ ($|h|$=0.004-0.45
for $\varepsilon$) means that a single power-law of
$P$$\propto$$h^{1/3}$ can robustly explain the experimental data,
even for $H$ very close to $H_{\rm c}$. Furthermore, our attempts
to fit the data at temperatures above 1.5 K reduce the linearity
range for both $\varepsilon$ and $P$ fits so that the power-law
becomes most evident at the lowest temperature available, $T$=0.66
K, in our experiment.

We find that the peculiar power-law of $P$ and $\varepsilon$ near
$T^*$ can be effectively described by a Landau-Ginzburg-type free
energy $F$=$F_{\rm ME}$+$F_{\rm E}$, where
\begin{equation}\label{eq2}
F_{\rm ME}=a{(\bf S^{{4+}} \cdot S^{3+})}P=a(H-H_{c})P
\end{equation}
and
\begin{equation}\label{eq3}
F_{\rm E}=\frac{b}{2}(T-T^*){^n}P^2+\frac{u}{4}P^4.
\end{equation}
The magnetoelectric free energy $F_{\rm ME}$ consists of a
coupling term between the spin orientations and the electric
polarization, while the ferroelectric free energy $F_{\rm E}$
effectively describes the evolution of $P$ as a crossover at
$T$$>$$T^*$ and as a 1st-order transition at $T$$<$$T^*$ under the
assumption that the coefficient of $P^2$ changes its sign across
$T^*$. Here, \emph{n} refers to an arbitrary exponent to describe
temperature-dependence of physical quantities near $T^*$ and
\emph{n}=1 in the classical case. The second-derivative of $F_{\rm
E}$ with respect to $P$ determines the dielectric susceptibility
$\chi^{-1}_{\rm e}$($P$)=$b$($T$-$T^*$)${^n}$+3$uP^{2}$, where
$\varepsilon_{\rm}$=$\chi_{\rm e}$($P$)+1. The sharp increase in
$\chi_{\rm e}$ observed in experiments can be understood as the
closeness of the system to the critical end point $T^*$. In our
Landau-Ginzburg model, when the temperature is approximately
$T^*$, the physics is driven by the interplay between the dominant
quartic $P^4$ term and linear $P$ term. Under this assumption,
$\partial$($F_{\rm ME}$+$F_{\rm E}$)/$\partial P$$\mid_{E=0}$=0
provides $P$$\propto$($H$-$H_{\rm c}$)$^{1/3}$. Furthermore,
$\partial^2$($F_{\rm ME}$+$F_E$)/$\partial
P^{2}$$\equiv$$\chi_{\rm e}^{-1}$=3$uP^{2}$$\propto$($H$-$H_{\rm
c}$)$^{2/3}$. These predicted exponents are consistent with those
determined experimentally (Figs. 3\textit{a} and 3\textit{b}),
clearly supporting the existence of $T^*$ near 0.66 K.

\begin{figure}
\begin{center}
\includegraphics[width=.46\textwidth]{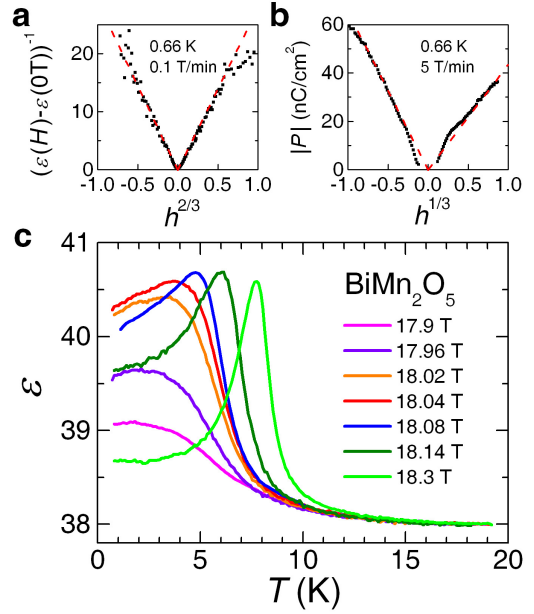}
\end{center}
\caption{Field-induced power-law evolution and $\varepsilon$
anomaly at very low temperature. (\textit{a})
($\varepsilon$(\textit{H})-$\varepsilon$(\textit{H}=0))$^{-1}$ vs.
$h^{2/3}$, where \textit{h}$\equiv$(\textit{H}-\textit{H}$_{\rm
c}$)/\textit{H}$_{\rm c}$ and $\mu_{0}$\textit{H}$_{\rm c}$=18.04
T. A very slow ramping rate (0.1 T/min) was used for the
$\varepsilon$(\textit{H}) measurement. The red dashed lines are
linear guides passing through the origin. The linear relationship
was valid in a wide $|h|$=0.004-0.45 (i.e., $\mu_{0}$$|H$-$H_{\rm
c}|$=0.1-8 T). (\textit{b}) Plot of $|P|$ vs. $h^{1/3}$. The
linear regime of $h$ was $|h|$=0.02-0.45 ($\mu_{0}$$|H$-$H_{\rm
c}$$|$=0.3-8 T). To increase the signal-to-noise ratio, it was
necessary to choose a rather fast ramping rate (5 T/min) for the
\textit{P}(\textit{H}) measurement, which made the \textit{P} data
less reliable in a small \textit{h} region. (\textit{c}) Detailed
$\varepsilon$(\textit{T}) curves down to \textit{T}$\sim$0.6 K at
fixed \textit{H} near \textit{H}$_{\rm c}$. The shape of the
$\varepsilon$(\textit{T}) curves above the maximum temperature was
similar to that predicted by Barrett's theory (Fig. S2).}
\label{fig3}
\end{figure}

The low value of $T^*$ found in our experiments suggests that
BiMn$_2$O$_5$ lies close to a a metaelectric quantum critical
point, where $T^{*}$ is actually zero. Closely analogous behavior
is well-known in the case of the itinerant metamagnet
Sr$_3$Ru$_2$O$_7$ \cite{grigera1}. In this respect, several novel
features of the present data particularly prompt further notice.
First, the observation of the robust power-law of
$P$$\propto$$h^{1/3}$ suggests a unique mean-field type while the
system is expected to have a 3D Ising universality class
($P$$\propto$$h^{1/4.82}$) near the classical critical end point
as seen in the liquid-gas transition. Moreover, the trajectory of
the crossover temperature near the zero temperature limit
following ($H$-$H_{\rm c}$)$^{1/2}$(Fig. 1\textit{e}) again
constitutes another mean-field type behavior. Thus, all these
observations suggest that the effective dimension of the system
has been lifted up to 4 or higher so as not to produce the
expected 3D fluctuation. One possible origin for such a
dimensional increase is in fact the dynamic exponent effect
expected due to the prevailing quantum critical end point
fluctuation.

The insights gained by our measurements and analysis above suggest
that the multiferroic BiMn$_2$O$_5$ provides a rare opportunity to
find a multi-ferroic quantum critical point and to test its
influence on ferroelectric materials. Supporting this novel
conclusion, detailed $\varepsilon$($T$) shape realized near
$H_{\rm c}$ (Fig. 3\textit{c}) further reveals a clue toward
detecting the quantum critical signature. When $H$ approaches
18.04 T from below, systematic increase of $\varepsilon$($T$=0.66
K) occurs. In particular, at $\mu_{0}H_{\rm c}$=18.04 T,
$\varepsilon$ continues to increase as $T$ decreases from 20 K and
shows a broad maximum around 4 K, followed by a slight decrease of
0.5 $\%$ down to $\sim$0.6 K. The behavior qualitatively resembles
that of the $\varepsilon$($T$) curve in quantum paraelectric
systems, such as SrTiO$_3$ \cite{dima,kvyat,wang,horiuchi}.
Furthermore, the increase in $\varepsilon$($T$) between 4 and 20 K
is not proportional to the inverse of temperature, as expected in
classical paraelectrics. Instead, it is closer to the shape
predicted by Barrett's theory (Fig. S2) \cite{barrett}, indicating
that the enhanced lattice fluctuation in close vicinity to $T^*$
is linked to \emph{the quantum mechanical population of the
relevant phonon mode.} Therefore, the physics of BiMn$_2$O$_5$
appears to involve quantum zero point lattice motion at low
temperatures as does SrTiO$_3$. The uniqueness of BiMn$_2$O$_5$
lies in the magnetic-field-tunability of its quantum-paraelectric
properties.

Our observations raises the interesting possibility that
instabilities into new quantum states of matter might develop in
close vicinity to the multi-ferroic critical end point, as they do
close to the magnetic quantum critical end point in
Sr$_3$Ru$_2$O$_7$. In future work we plan to investigate this
possibility by fine-tuning the $T^*$ position, either by by
rotating the field direction or by applying external pressure.

\section*{Methods}
Single crystal samples were grown using the flux method
\cite{hur1}. All the electrical properties were measured along the
\textit{b}-axis while the magnetic field was applied along the
\textit{a}-axis up to 35 T at the National High Magnetic Field
Laboratory (NHMFL). Dielectric constants were measured by a
capacitance bridge (GR1615A or AH2500A) at 1 kHz \cite{haam}.
\textit{P} was obtained by integration of the pyro- and
magnetoelectric currents measured by use of an electrometer
(Keithley 617). Given the limited magnet time for high-field
measurements, the continuous application of a small electric field
bias of \textit{E}=2 kV/cm was quite useful in quickly determining
the $P$ evolution. In particular, the value of \textit{E}=2 kV/cm
was sufficient to fully polarize the sample on cooling through
\textit{T}$_{\rm C}$ but small enough to produce a \textit{P}
close to its spontaneous value. In several field and temperature
sweeps, it was verified that the spontaneous \textit{P} measured
under zero electric field after poling with \textit{E}=2 kV/cm is
close to the \textit{P} measured under the same bias. A
precalibrated capacitance sensor was used for accurate
\textit{T}-sweeps between 0.6 and 20 K near \textit{H}$_{\rm c}$
(Fig. 3\textit{c}), while a Cernox sensor was mostly used for the
\textit{T}- and \textit{H}-sweeps above 4 K. For \textit{H}-swept
measurements at \textit{T}=0.66 K and 1.5 K, temperatures were
stabilized by controlling the vapor pressure. Specific heat and
magnetocaloric effects were measured with a plastic calorimeter.
Temperature-dependent DC magnetization was measured using a
vibrating sample magnetometer in a Physical Property Measurement
System (Quantum Design). A pickup-coil magnetometer was used to
measure the \textit{H}-dependent magnetization in a pulse magnet
at NHMFL-Los Alamos National Laboratory.

\begin{acknowledgments}
We thank P. Chandra, S. Rowley, J. Schmalian, and G. R. Stewart
for stimulating discussions. This study was supported by Korean
Government through NRL (M10600000238) and GPP programs, and by KRF
(KRF-2008-205-C00101). JW was supported by Seoul R\&BD. Work at
Rutgers was supported by NSF-0405682 and NSF-DMR-0605935. Work at
NHMFL was performed under the auspices of the National Science
Foundation, the State of Florida, and the U. S. Department of
Energy.
\end{acknowledgments}


\begin{thebibliography}{99}

\bibitem[*]{khkim} Corresponding author: khkim@phya.snu.ac.kr

\bibitem{grigera1} Grigera SA \emph{et al.} (2001) Magnetic
field-tuned quantum criticality in the metallic ruthenate
Sr$_{3}$Ru$_{2}$O$_{7}$. \emph{Science} 294:329-332.

\bibitem{neil} Harrison N, Jaime M, Mydosh JA (2003) Reentrant
hidden order at a metamagnetic quantum critical end point.
\emph{Phys Rev Lett} 90:096402.

\bibitem{grigera2} Grigera SA \emph{et al.} (2004) Disorder-sensitive
phase formation linked to a metamagnetic quantum criticality.
\emph{Science} 306:1154-1157.

\bibitem{khkim1} Kim KH \emph{et al.} (2004) Nexus between quantum critically and phase formation in
U(Ru$_{1-x}$Rh$_{x}$)$_{2}$Si$_{2}$. \emph{Phys Rev Lett}
93:206402.

\bibitem{Kutnjak} Kutnjak Z, Petzelt J, Blinc R (2006) The giant
electromechanical response in ferroelectric relaxors as a critical
phenomenon. \emph{Nature}, 441:956-959.

\bibitem{vekhter} Vekhter BG, Kaplan MD (1979) A new type of structural Jahn-Teller phase
transition in DyVO$_{4}$ induced by an electric field. \emph{JETP
Lett} 29:155-157.


\bibitem{kimura} Kimura T \emph{et al.} (2003) Magnetic control of ferroelectric polarization.
 \emph{Nature} 426:55-58.

\bibitem{hur1} Hur N \emph{et al.} (2004) Electric polarization
reversal and memory in a multiferroic material induced by magnetic
fields. \emph{Nature} 429:392-395.

\bibitem{spaldin} Spaldin NA, Fiebig M (2005) The
renaissance of magnetoelectric multiferroics. \emph{Science}
309:391-392.

\bibitem{eerenstein} Eerenstein W, Mathur ND, Scott JF (2006) Multiferroic and magnetoelectric
 materials. \emph{Nature} 442:759-765.

\bibitem{tokura} Tokura Y (2006) Multiferroics as
quantum electromagnets. \emph{Science} 312:1481-1482.

\bibitem{kenzelmann} Kenzelmann M \emph{et al.} (2007) Direct transition from a disordered to
a multiferroic phase on a triangular lattice. \emph{Phys Rev Lett}
98:267205.

\bibitem{munoz} Mu\~{n}oz A \emph{et al.} (2002) Magnetic
structure and properties of BiMn$_{2}$O$_{5}$ oxide: A neutron
diffraction study. \emph{Phys Rev B} 65:144423.

\bibitem{vec} Vecchini C \emph{et al.} (2008) Commensurate magnetic structures of
\emph{R}Mn$_{2}$O$_{5}$ (\emph{R}=Y, Ho, Bi) determined by
single-crystal neutron diffraction. \emph{Phys Rev B} 77:134434.

\bibitem{katsura} Katsura H, Nagaosa N, Balatsky AV (2005) Spin
current and magnetoelectric effect in noncollinear magnets.
\emph{Phys Rev Lett} 95:057205.

\bibitem{chapon1} Chapon LC \emph{et al.} (2004) Structural
anomalies and multiferroic behavior in magnetically frustrated
TbMn$_{2}$O$_{5}$. \emph{Phys Rev Lett} 93:177402.

\bibitem{chapon2} Chapon LC \emph{et al.} (2006)
Ferroelectricity induced by acentric spin-density waves in
YMn$_{2}$O$_{5}$. \emph{Phys Rev Lett} 96:097601.

\bibitem{cheong} Cheong S-W, Mostovoy M (2007) Multiferroics: a magnetic twist for ferroelectricity. \emph{Nat
Mater} 6:13-20.

\bibitem{blazey} Blazey KW, Rohrer H, Webster R (1971) Magnetocaloric effects and the angular variation of the magnetic
phase diagram of antiferromagntic GdAlO$_{3}$. \emph{Phys Rev B}
4:2287-2302.

\bibitem{king} King AR, Rohrer H (1979) Spin-flop bicritical field in MnF$_{2}$. \emph{Phys Rev B} 19:5864-5876.

\bibitem{kanamori} Kanamori J (1963) in \emph{Magnetism}, eds Rado GT \& Suhl H (Academic Press, New York), pp 127-203.

\bibitem{koo} Koo J \emph{et al.} (2007) Non-resonant and resonant
X-ray scattering studies on multiferroic TbMn$_{2}$O$_{5}$.
\emph{Phys Rev Lett} 99:197601.

\bibitem{haam} Haam SY \emph{et al.} (2006) Evolution of
ferroelectric and antiferromagnetic phases of TbMn$_{2}$O$_{5}$
under high magnetic field up to 45 T. \emph{Ferroelectrics}
336:153-159.

\bibitem{stryj} Stryjewski E, Giordano N (1977) Metamagnetism.
\emph{Adv Phys} 26:487-650.

\bibitem{dima} Khmel'nitskii DE, Shneerson VL (1971) Low-temperature displacement-type phase
transition in crystals. \emph{Sov Phys Solid State} 13:687-694.

\bibitem{kvyat} Kvyatkovski\u{i} OE (2001) Quantum effects
in incipient and low-temperature ferroelectrics (a review).
\emph{Phys Solid State} 43:1401-1419.

\bibitem{wang} Wang R, Sakamoto N, Itho M (2000) Effects of pressure on the dielectric properties of SrTi$^{18}$O$_3$ and
SrTi$^{16}$O$_{3}$ single crystals. \emph{Phys Rev B}
62:R3577-R3580.

\bibitem{horiuchi} Horiuchi S, Okimoto Y, Kumai R, Tokura Y (2003) Quantum phase transition in
organic charge-transfer complexes. \emph{Science} 299:229-232.

\bibitem{barrett} Barrett JH (1952) Dielectric constant in perovskite type crystals. \emph{Phys Rev} 86:118-120.

\end{thebibliography}
\end{document}